\documentclass[conference]{IEEEtran}
\IEEEoverridecommandlockouts
\usepackage{Style/physicsarticle}

\begin{document}
\bstctlcite{IEEEexample:BSTcontrol}
\title{Fast quantum gate design with deep reinforcement learning using real-time feedback on readout signals\\
\thanks{This work was supported by the Natural Sciences and Engineering Research Council of Canada (NSERC) through its Discovery (RGPIN-2020-04328), CREATE (543245-2020), and CGS M programs. Cette oeuvre a \'{e}t\'{e} soutenue par le Conseil de recherches en sciences naturelles et en génie du Canada (CRSNG) via ses programmes Discovery (RGPIN-2020-04328), CREATE (543245-2020) et CGS M.
%
}
}

\author{\IEEEauthorblockN{Emily Wright}
\IEEEauthorblockA{\textit{Department of Physics and Astronomy} \\
\textit{University of Victoria}\\
Victoria, Canada \\
emilywright1@uvic.ca}
\and
\IEEEauthorblockN{Rog\'{e}rio de Sousa}
\IEEEauthorblockA{\textit{Department of Physics and Astronomy} \\
\textit{University of Victoria}\\
Victoria, Canada \\
rdesousa@uvic.ca}
}

\maketitle

\begin{abstract}
The design of high-fidelity quantum gates is difficult because it requires the optimization of two competing effects, namely maximizing gate speed and minimizing leakage out of the qubit subspace. 
We propose a deep reinforcement learning algorithm that uses two agents to address the speed and leakage challenges simultaneously. 
The first agent constructs the qubit in-phase control pulse using a policy learned from rewards that compensate short gate times.
The rewards are obtained at intermediate time steps throughout the construction of a full-length pulse, allowing the agent to explore the landscape of shorter pulses.
The second agent determines an out-of-phase pulse to target leakage. 
Both agents are trained on real-time data from noisy hardware, thus providing model-free gate design that adapts to unpredictable hardware noise.
To reduce the effect of measurement classification errors, the agents are trained directly on the readout signal from probing the qubit.
We present proof-of-concept experiments by designing X and square root of X gates of various durations on IBM hardware.
After just 200 training iterations, our algorithm is able to construct novel control pulses up to two times faster than the default IBM gates, while matching their performance in terms of state fidelity and leakage rate.
As the length of our custom control pulses increases, they begin to outperform the default gates.
Improvements to the speed and fidelity of gate operations open the way for higher circuit depth in quantum simulation, quantum chemistry and other algorithms on near-term and future quantum devices.
\end{abstract}


\date{\today}

\maketitle

\newpage

\section{Introduction}
\label{sec:intro}
Hardware noise, fabrication variability, and imperfect logic gates are the greatest barriers to performing reliable quantum computations at a large-scale~\cite{Preskill2018}.
Presently, gate operations on quantum computers such as those produced by IBM are realized with derivative removal by adiabatic gate (DRAG) pulses calculated analytically from a simple three-level model~\cite{Motzoi2009}. 
The fidelity of these gate operations suffers due to long gate times, imperfect models, and time-dependent changes in the processor parameters such as qubit frequencies.
Frequent calibration to combat these fluctuations is costly and even when properly calibrated, the control pulse shapes are sub-optimal and allow for errors.
Decoherence and leakage out of the computational sub-space are of particular concern in the context of fault-tolerant quantum computing as they require substantial additional resources to correct and can significantly impact the threshold of certain error correction codes~\cite{McEwen2021, Wood2018, Suchara2014, Fowler2013}.
Thus, engineering faster and higher-fidelity gates is of timely importance. 

Existing strategies for gateset design are analytic~\cite{Khaneja2001, Boscain2002, Pechen2006, Montangero2007, Motzoi2009} or based on numerical simulations that require precise physical and noise models of the hardware~\cite{Krotov1993, Dehaghani2022, Zhu1998, DAllessandro2001, Khaneja2005, Machnes2018, Doria2011}.
In large-scale quantum processors, the difficulty of completely characterizing the system prohibits model-based control techniques.
Reinforcement learning (RL)~\cite{Tsitsiklis1994, Watkins1992, Szepesvari1999} is an alternative approach for gate design which operates without prior knowledge of the hardware model.
RL and its variants have been applied to myriad quantum control problems using numerically simulated environments~\cite{Metz2022, Qiu2022, Peng2021, An2021, Daraeizadeh2020, Ma2020, Chen2019, Porotti2019, Zhang2019, Chen2014, Shindi2023, Sivak2022, Guo2021, Kuo2021, Wauters2020, Garciasaez2019, Bukov2018}.
Such set-ups demonstrate the potential of RL, but suffer from the same modelling constraints as other optimization methods.
Recently, a few experiments have been carried out using RL directly on noisy hardware~\cite{Reuer2022, Baum2021, Borah2021, Bukov2018_2}. 

In this paper, we propose a new RL algorithm to design fast quantum gates.
Our algorithm has several advantages over existing proposals including:
\begin{enumerate}
    \item enabling design of faster gates by rewarding intermediate steps in the control pulse,
    \item reducing the impact of measurement errors by training directly on the readout signal rather than classifying the state,
    \item mitigating leakage with a dual agent architecture, 
    \item speeding up training using low measurement overhead and real-time feedback, and
    \item accounting for realistic noise in the quantum processor by training directly on hardware.
\end{enumerate}
As well, we initialize the agent by pre-training on a calibrated DRAG pulse to capture information about the system dynamics.
We optimize $X$ and $\sqrt{X}$ gates of different durations as a proof-of-concept; however, our algorithm can easily be extended to two-qubit gates by modifying the reward function and state space to complete a universal gateset.

The paper is structured as follows: in Section \ref{sec:RL} we introduce the general RL algorithm, in Section \ref{sec:lit} we review the related literature, in Section \ref{sec:algo} we describe our deep RL algorithm for fast quantum gate design, in Section \ref{sec:exp} we present experimental results, and in Section \ref{sec:gen} we describe the extension of our algorithm to two-qubit gates.

\begin{figure*}
\centering
\includegraphics[clip,width=0.8\textwidth]{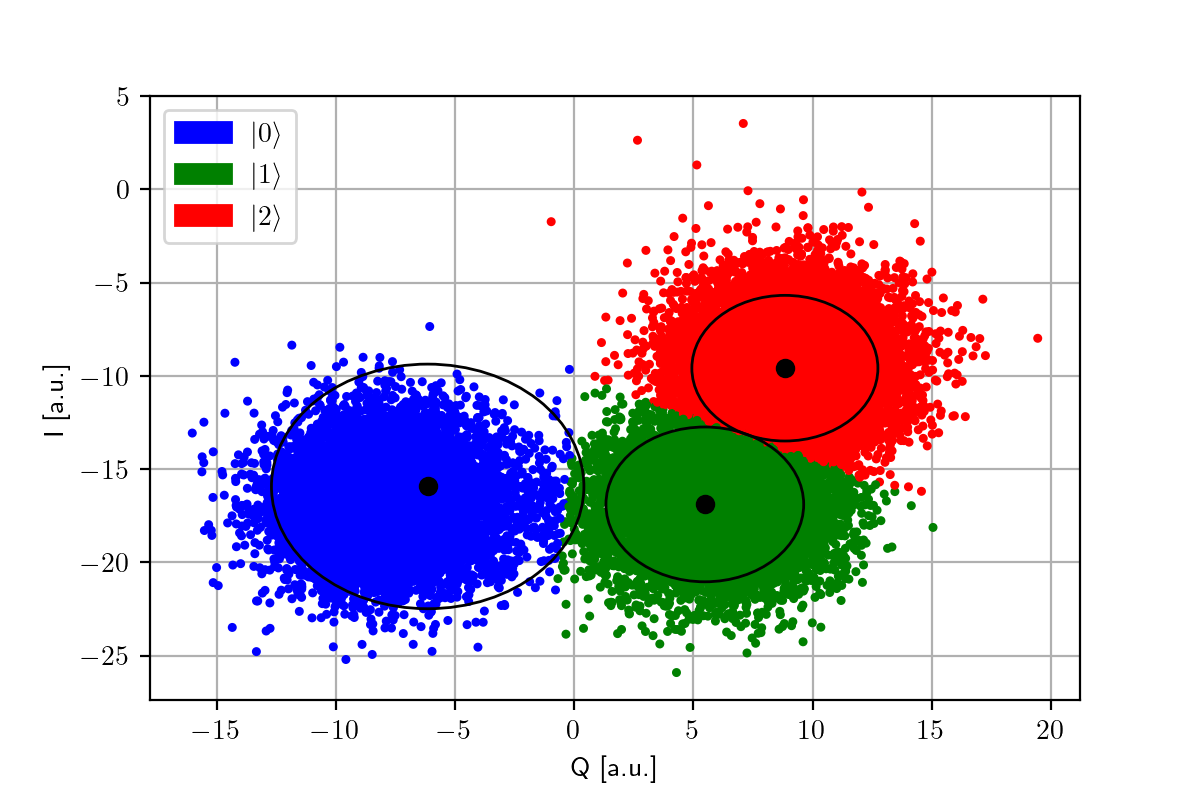}
\caption{A plot of the readout signal for different qubit states (10000 shots each) on IBM Lima. Black circles indicate the mean and standard deviation of each cluster.
The overlap shows the difficulty of distinguishing between states.}
\label{fig:stateData}
\vspace{-6mm}
\end{figure*}

\section{The RL algorithm}
\label{sec:RL}
In this section, we describe the RL algorithm in detail.
RL is a machine learning algorithm wherein an agent interacts with a system and iteratively updates a control policy based on its observations.
The entire process is modelled as a controlled Markov Decision Process (MDP).
Let $\bb{S} \subset \bb{R}^n$ be the space of states for the system and $\bb{U}$ the set of possible actions.
At each step, the system is in some state $s_j$ and the agent decides on an action $u_j$ according to a policy. 
The policy $\pi(\cdot|s_j)$ is a conditional probability distribution over the possible actions in $\bb{U}$ given the current state.
The system moves into the next state $s_{j+1}$ via a stochastic transition kernel $\T(\cdot|s_j,u_j)$. 
After observing the next state, the agent receives a corresponding reward $r_j(s_j, u_j, s_{j+1})$.
The objective of the controller is to maximize the infinite-horizon discounted expected reward
\begin{equation}
    \label{eq:QLReward}
    J_\beta(s_0, \pi) = E_{s_0}^{\T, \pi} \left[\sum_{j=0}^\infty \beta^j r_j(s_j, u_j, s_{j+1}) \right]
\end{equation}
over the set of admissible policies $\pi$, where $0 < \beta < 1$ is a discount factor and $E_{s_0}^{\T, \pi}$ denotes the expectation for initial state $s_0$ and transition kernel $\T$ under policy $\pi$.

The standard RL algorithm is Q-learning~\cite{Watkins1992}. 
Q-learning involves tracking the ``value'' of taking an action $u_j$ given the current state $s_j$.
The Q-value is stored in a table indexed by states and actions.
Given an initial table $Q_0$, the value of each state-action pair is updated according to the Bellman equation 
\begin{align}
    \label{eq:bell}
    Q_{j+1}(s_j, u_j) &= Q_j(s_j, u_j) +
    \alpha_j(s_j, u_j) \Big[r(s_j, u_j, s_{j+1}) \\
    &+ \beta \max_{v \in \bb{U}} Q_j(s_{j+1}, v) - Q_j(s_j, u_j)\Big] \notag
\end{align}
as the agent explores the environment~\cite{Watkins1992}.
The coefficient $\alpha_j$ is a hyperparameter called the learning rate and determines how quickly the agent adapts to changes in the environment.
Under mild conditions on the learning rate, the algorithm converges to a fixed point denoted $Q_*$, which satisfies
\begin{equation}
    \label{eq:fixpt}
    Q_*(s, u) = E\left[r(s, u, s') + \beta \max_{v\in\bb{U}} Q_*(s', v)\Big| s, u\right].
\end{equation}
A policy $\pi$ which satisfies 
\begin{equation}
    \max_{u\in\bb{U}} Q_*(s, u) = Q_*(s, \pi(s))
\end{equation}
is an optimal policy (see Theorem 4 in~\cite{Tsitsiklis1994} and the main Theorem in~\cite{Watkins1992}).

The Q-learning algorithm was conceived for finite action and state spaces. 
For large and/or continuous state spaces, storing the Q-value in a table is not an option.
To overcome this challenge, one might quantize the state space~\cite{Gray1998} or use function approximation~\cite{Gadjiev2010}.
While Q-learning with quantization or function approximation is not guaranteed to converge, there is ample empirical evidence that it can be used to solve quantum control problems~\cite{Metz2022, Qiu2022, Peng2021, An2021, Daraeizadeh2020, Ma2020, Chen2019, Porotti2019, Zhang2019, Chen2014, Shindi2023, Sivak2022, Guo2021, Kuo2021, Wauters2020, Garciasaez2019, Bukov2018, Reuer2022, Baum2021, Borah2021, Bukov2018_2}.
In this work, we approximate the Q-table using a neural network, in a strategy that has been termed ``deep reinforcement learning'' (DRL)~\cite{Bertsekas1996, Li2018}.
The state $s_j$ is the input to the neural network and the output is a probability distribution $\pi(\cdot|s_j)$ over the action space $\bb{U}$.
The neural network is represented by a set of parameters $\theta$ which are updated in a manner that approximates the Bellman equation~\cite{Williams1992}.

\section{RL for quantum gate design}
\label{sec:lit}
Having introduced RL, we now review its uses for quantum gate design to-date.
Many RL algorithms have been proposed to solve quantum control problems in areas ranging from Hamiltonian engineering~\cite{Peng2021} to quantum metrology~\cite{Qiu2022}.
Theoretical algorithms make use of simulated environments to provide full access to the state of the quantum system~\cite{Metz2022, Qiu2022, Peng2021, An2021, Daraeizadeh2020, Ma2020, Chen2019, Porotti2019, Zhang2019, Chen2014}.
Of these proposals, several specifically target unitary gate design~\cite{An2019, Niu2019, Daraeizadeh2020}.
In all cases, the agent has access to the exact unitary operator specifying the Schr\"{o}dinger evolution of the system.
In~\cite{An2019, Daraeizadeh2020} a simplified Hamiltonian model is used while~\cite{Niu2019} simulates a gmon environment which mimics noisy control actuation and incorporates leakage errors.
The agents receive rewards based on gate infidelity which is inaccessible in experiments.
In simulation, these algorithms are able to achieve improvements in gate fidelity~\cite{An2019, Niu2019, Daraeizadeh2020} and gate time~\cite{Niu2019, Daraeizadeh2020} over other gate synthesis strategies.
However, these proposals are not compatible with training on real hardware thus necessarily suffer from model bias.
More realistic RL set-ups for quantum control only provide access to fidelities and/or expectation values for some observables~\cite{Shindi2023, Sivak2022, Guo2021, Kuo2021, Wauters2020, Garciasaez2019, Bukov2018}.
Specifically for gate design, Shindi et al. proposed in~\cite{Shindi2023} to probe a gate with a series of input states.
The reward incorporates the average fidelity between the output states and the target states.
Such algorithms require prohibitive amounts of averaging in experiments so they are also confined to numerical simulations.

The next step is to perform RL using stochastic measurement outcomes or low-sample estimators of physical observable~\cite{Reuer2022, Baum2021, Borah2021, Bukov2018_2}.
Most recently, some experiments have been carried out using RL directly on noisy quantum hardware~\cite{Reuer2022, Baum2021}. 
Baum et al. trained a DRL agent on a superconducting computer for error-robust gateset design~\cite{Baum2021}.
The agent was able to learn novel pulse shapes up to three times faster than industry standard DRAG gates with slightly lower error per gate and improvements maintained without re-calibration for up to 25 days.
The search space was restricted to 8 and 10 segment piece-wise constant operations for one- and two-qubit gates respectively.
Despite the small search space, the optimization algorithm was inefficient. 
The training was completed in batches and the reward was a weighted mean over fidelities estimated using full state tomography.
Subsequently, Reuer et al. developed a latency-based DRL agent implemented via an FPGA capable of using real-time feedback at the microsecond time scale~\cite{Reuer2022}. 
They demonstrated its effectiveness with a state preparation experiment on a superconducting qubit.
In this paper, we look ahead to a future where real-time feedback for quantum control is common.
We improve upon the work in~\cite{Baum2021} by allowing greater flexibility in pulse shapes, training directly on the readout signal to reduce the impact of measurement errors, specifically targeting leakage errors, and relying on real-time feedback to speed-up the learning process.

\section{Our DRL algorithm for fast quantum gate design}
\label{sec:algo}
We now describe our DRL algorithm for design of fast quantum gates in detail.
Here, we outline our algorithm specifically for superconducting transmon qubits, but it can easily be adapted to other hardware platforms such as trapped ions~\cite{Bruzewicz2019}, quantum dots~\cite{Vandersypen2019}, or neutral atoms~\cite{Henriet2020}. 
In this case, the quantum system consists of a transmon qubit dispersively coupled to a superconducting resonator and controlled via a capacitively coupled voltage drive line (see Appendix \ref{sec:hardware} for details).
The voltage drive envelope 
\begin{equation}
    \label{eq:drive}
    c(t) = \begin{cases}
    c^x(t)\cos(\omega_dt) + c^y(t)\sin(\omega_dt) &0 < t < t_g \\
    0 &\text{otherwise}
    \end{cases}
\end{equation}
is composed of two independent quadrature controls $c^{x}(t)$ and $c^{y}(t)$ on a single drive frequency $\omega_d$ for the duration of the gate operation $t_g$.
We seek to design a piece-wise constant (PWC) control pulse with up to $N_\textup{seg}$ segments of equal length $\tau = t_g / N_\textup{seg}$ where $t_g$ is the maximum duration of the gate.
 
Our DRL algorithm consists of two neural networks to decide sequentially on the $x$- and $y$-quadrature of the control pulse.
The dual agent structure enables us to mitigate leakage errors while simultaneously creating faster gates. 
At each time step, our agents select the amplitudes of the next segment based on real-time feedback about the state of the system.
The agents receive rewards throughout the construction of the pulse (rather than just at the end) which opens the possibility to design faster gates.
We eliminate model-bias by training on quantum hardware so our DRL algorithm can account for realistic noise in the qubits and for other errors such as over-rotation introduced by the classical drive lines.
We train directly on the observation signal resulting from probing the qubit.
The signal has two components which we denote $I$ for ``in-phase'' and $Q$ for ``quadrature''.
Previous algorithms for gate design have used the $\ket{0}$ and $\ket{1}$ populations imperfectly estimated from the location of the signal in the $(I,Q)$-plane.
Figure \ref{fig:stateData} shows readout data taken on the IBM Lima quantum computer, where there is overlap between the locations of the $\ket{0}$, $\ket{1}$, and $\ket{2}$ measurements.
Our algorithm reduces the impact of measurement errors because we avoid classifying the signal into $\ket{0}$ or $\ket{1}$.
It also has a low measurement overhead since we do not perform full state tomography.

The network parameters $\theta^x, \theta^y$ are updated periodically during the training.
Each training iteration for $i=0,\hdots,N_\textup{iter}-1$ consists of $N_\textup{ep} \leq N_\textup{seg}$ episodes counted using the index $j$.
The ratio $N_\textup{seg}/N_\textup{ep}$ is an integer that sets how many times the neural network parameters are updated before a full waveform is constructed.
We use a second index $k = k(i,j)$ to track the waveform segment.
The $j$-th input to the first neural network is a state $s_j^x = (\langle I_j\rangle, \langle Q_j\rangle, k)$ composed of the average $(I,Q)$ signal over $N_\text{shot}$ measurements and the current segment index $k$.
The output of the agent is a probability distribution over the action space $\bb{U}^x$ which we denote $\pi_{\theta^x_i}(\cdot|s_j^x)$ to indicate the dependence on the neural network parameters $\theta_i^x$.
The agent samples an action $u_j^x \in \bb{U}^x$ according to $\pi_{\theta^x_i}(\cdot|s_j^x)$.
The state for the second agent $s_j^y = (u_j^x, \Ell_j)$ is formed of the amplitude $u_j^x$ on the first quadrature and the leakage population
\begin{equation}
    \label{eq:leak}
    \Ell_j = \left|\bra{2}U_{k+1}\ket{0}\right|^2
\end{equation}
estimated from the $(I,Q)$-plane, where $U_{k+1}$ represents the operation of evolving the qubit by the first $k+1$ segments of the waveform.
The agent now samples an action $u_j^y \in \bb{U}^y$ according to the output $\pi_{\theta^y_i}(\cdot|s_j^y)$ of the second network.
The $k$-th segment of the control pulse thus takes the form
\begin{equation}
    c(t) = u_j^x \cos(\omega_dt) + u_j^y \sin(\omega_dt)
\end{equation}
for $k \tau \leq t < (k+1)\tau$.
Based on hardware parameters, we restrict the $x$-quadrature amplitudes to the set $\bb{U}^x = \{0.00, 0.01, \hdots, 0.19, 0.20\}$ and the $y$-quadrature amplitudes to $\bb{U}^y = \{-0.10, -0.09, \hdots, 0.09, 0.10\}$.

To train the agent, we require a reward function for each network.
Let $c_T = (\langle I_T\rangle, \langle Q_T\rangle) \pm \sigma_T$ be the expected average measurement result after applying the target gate to the ground state (calibrated experimentally).
During training, we penalize the distance of the observed measurement signal from $(\langle I_T\rangle, \langle Q_T\rangle)$ which encapsulates both a failure to steer the qubit into the desired state and leakage into higher energy levels. 
We use the reward function
\begin{align}
    \label{eq:rx}
    &r_j^x\left(s_{j+1}^x, c_T\right) =
    \\
    &\min \left\{1 - \lambda k, \frac{\sigma_T}{\|(\langle I_{j+1}\rangle,\langle Q_{j+1}\rangle) - (\langle I_T\rangle, \langle Q_T\rangle)\|}
         - \lambda k\right\} \notag
\end{align}
to train the first network, where the term $\lambda k$ penalizes the length of the control pulse for some coefficient $\lambda \in \bb{R}$.
For the second network, we compensate low leakage populations with the reward
\begin{equation}
    \label{eq:ry}
    r_j^y\left(s_{j+1}^y\right) = \max\left\{0, \Ell_\text{max} - \Ell_{j+1}\right\}
\end{equation}
where $\Ell_\text{max}$ sets a limit on the allowable leakage.
A summary of our DRL algorithm is shown in Algorithm \ref{alg:RL}.

\begin{algorithm}
\caption{DRL for fast quantum gate design}
\label{alg:RL}
\begin{algorithmic}[1]
\REQUIRE initial state $s_0^x$, leakage $\Ell_0$, and parameters $\theta_0^x, \theta_0^y$
\STATE Set $k = 0$ 
\FOR{each iteration $i=0, ..., N_\textup{iter}-1$}
    \FOR{each episode $j=0, ..., N_\textup{ep}-1$}
        \IF{$k = N_\textup{seg}$} 
            \STATE Re-initialize waveform $k=0$
            \STATE Re-initialize state $s^x_j = s^x_0$
            \STATE Re-initialize leakage $\Ell_j = \Ell_0$
        \ENDIF
        \STATE Select next action $u^x_j$ according to policy $\pi_{\theta^x_i}(\cdot|s_j^x)$ 
        \STATE Set $s_j^y = (u_j^x, \Ell_j)$
        \STATE Select next action $u^y_j$ according to policy $\pi_{\theta^y_i}(\cdot|s_j^y)$ 
        \STATE Evolve qubit by first $k+1$ segments of waveform
        \STATE Measure qubit $N_\textup{shot}$ times to get $(\langle I_{j+1}\rangle, \langle Q_{j+1}\rangle)$ 
        \STATE Update $k \to k + 1$
        \STATE Set $s^x_{j+1} = (\langle I_{j+1}\rangle, \langle Q_{j+1}\rangle, k)$
        \STATE Calculate reward $r_j^x$
        \STATE Estimate leakage population $\Ell_{j+1}$
        \STATE Calculate reward $r_j^y$
        \STATE Reset qubit to $\ket{0}$
    \ENDFOR
    \STATE Send trajectory $\left(s_0^x, u_0^x, r_0^x, s_1^x, u_1^x, r_1^x, \hdots, r_{N_\textup{ep}-1}^x\right)$ to first network
    \STATE Update neural network parameters $\theta_{i}^x \to \theta_{i+1}^x$ 
    \STATE Send trajectory $\left(s_0^y, u_0^y, r_0^y, s_1^y, u_1^y, r_1^y, \hdots, r_{N_\textup{ep}-1}^y\right)$ to second network
    \STATE Update neural network parameters $\theta_{i}^y \to \theta_{i+1}^y$ 
\ENDFOR
\end{algorithmic}
\end{algorithm}

The initial state $s_0^x$ and leakage population $\Ell_0$ are estimated by measuring the qubit before performing any gate operations.
The initial policies $\pi_{\theta_0}^x, \pi_{\theta_0}^y$ are generated based on the industry standard DRAG gate by pre-training the agent.
The DRAG pulse is Gaussian with a derivative component on the second quadrature.
That is, $c^x(t) = \Omega_G(t)$ and $c^y(t) = \gamma \Omega_G'(t)$ for a Gaussian envelope $\Omega_G$ and a coefficient $\gamma \in \bb{R}$~\cite{Motzoi2009}.
The DRAG pulse is designed to reduce leakage based on a simple three-level model.
In experiments, the amplitude of $\Omega_G$ and the $\gamma$ factor are calibrated to combat time dependent changes in the noise.
Our initial policy captures information about the system dynamics using the calibrated DRAG pulse; however, our algorithm should not be considered an optimization of the DRAG pulse which has been proposed in other works~\cite{Motzoi2009, Werninghaus2021}.
In the next section, we describe experimental results showing that our agent learns novel pulse shapes able to outperform the DRAG gate in terms of fidelity and leakage rate.
For more details on the pre-training, see Appendix \ref{sec:pretrain}.


\begin{figure*}
    \centering
    \begin{subfigure}[b]{0.49\textwidth}
        \includegraphics[clip,width=\textwidth]{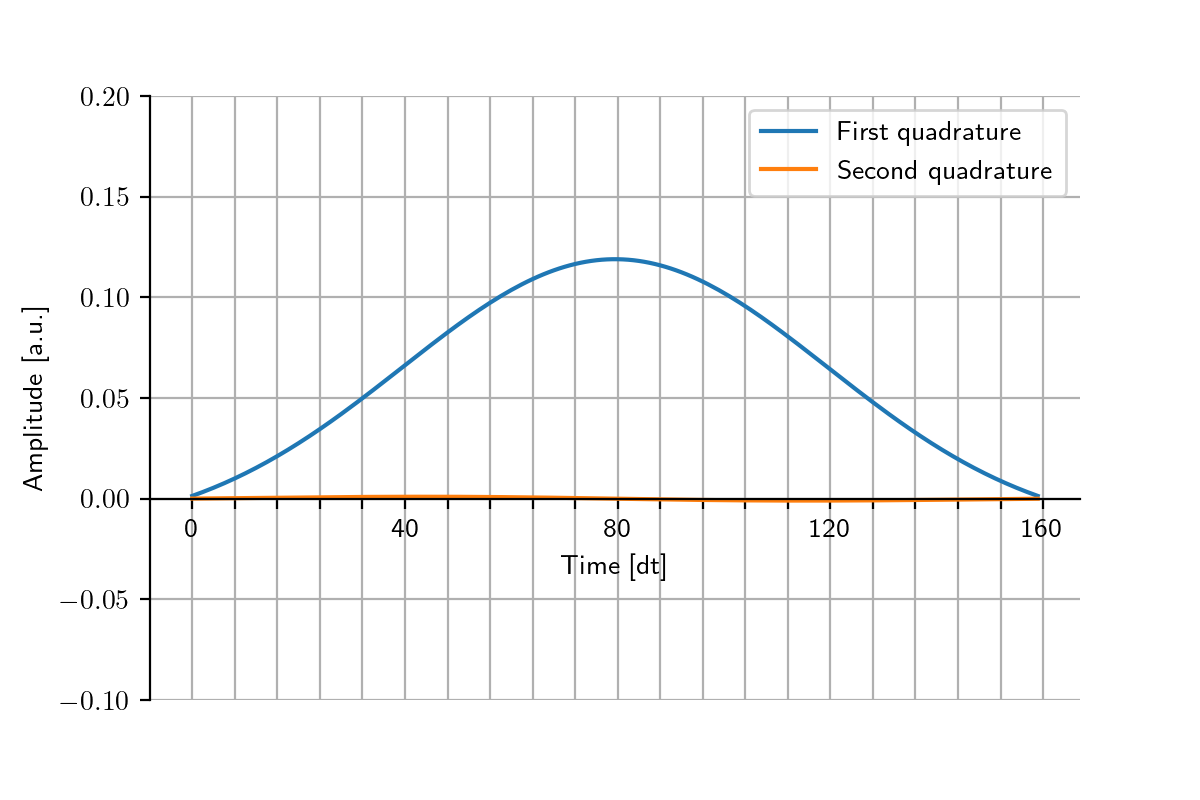}
        \vspace{-10mm}
        \caption{DRAG $X$ gate.}
        \label{fig:XresA}
    \end{subfigure}
    \hspace{-5mm}
    \begin{subfigure}[b]{0.49\textwidth}
        \includegraphics[clip,width=\textwidth]{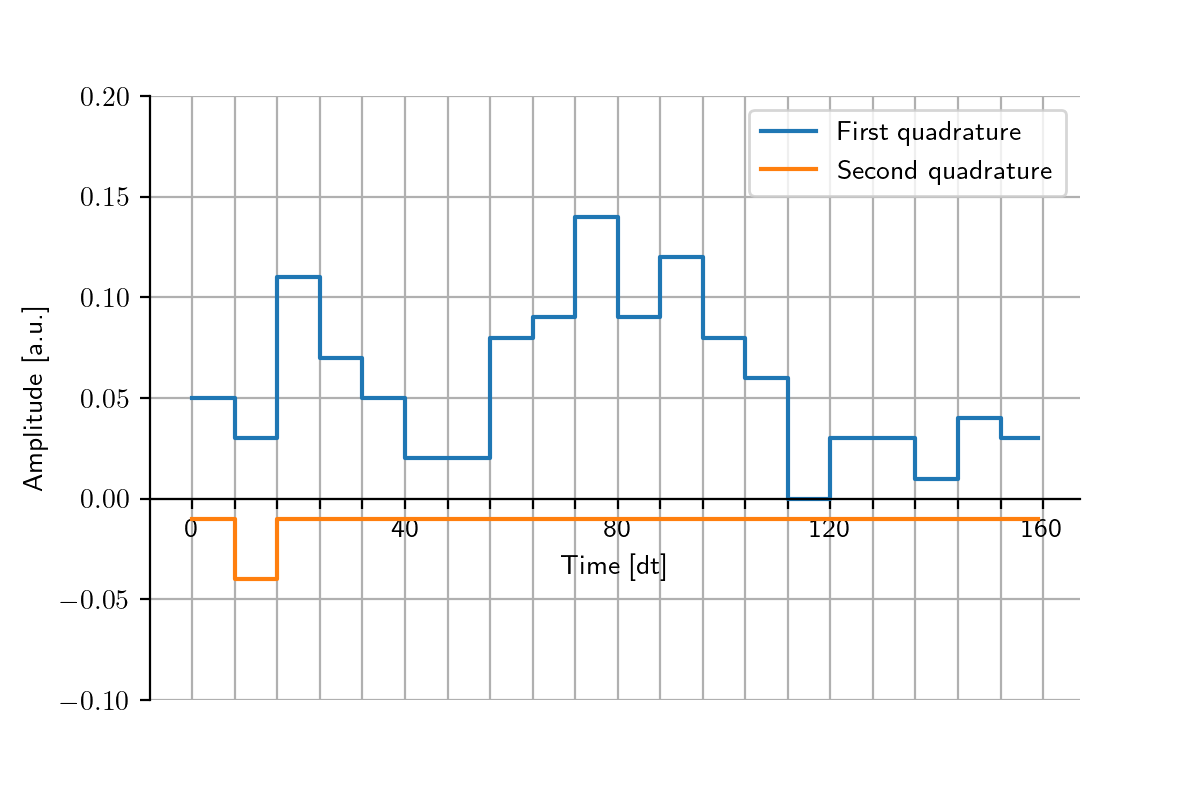}
        \vspace{-10mm}
        \caption{Optimized $X$ gate.}
        \label{fig:XresB}
    \end{subfigure}
    \vspace{-4mm}
    \\
    \centering
    \begin{subfigure}[b]{0.49\textwidth}
        \includegraphics[clip,width=\textwidth]{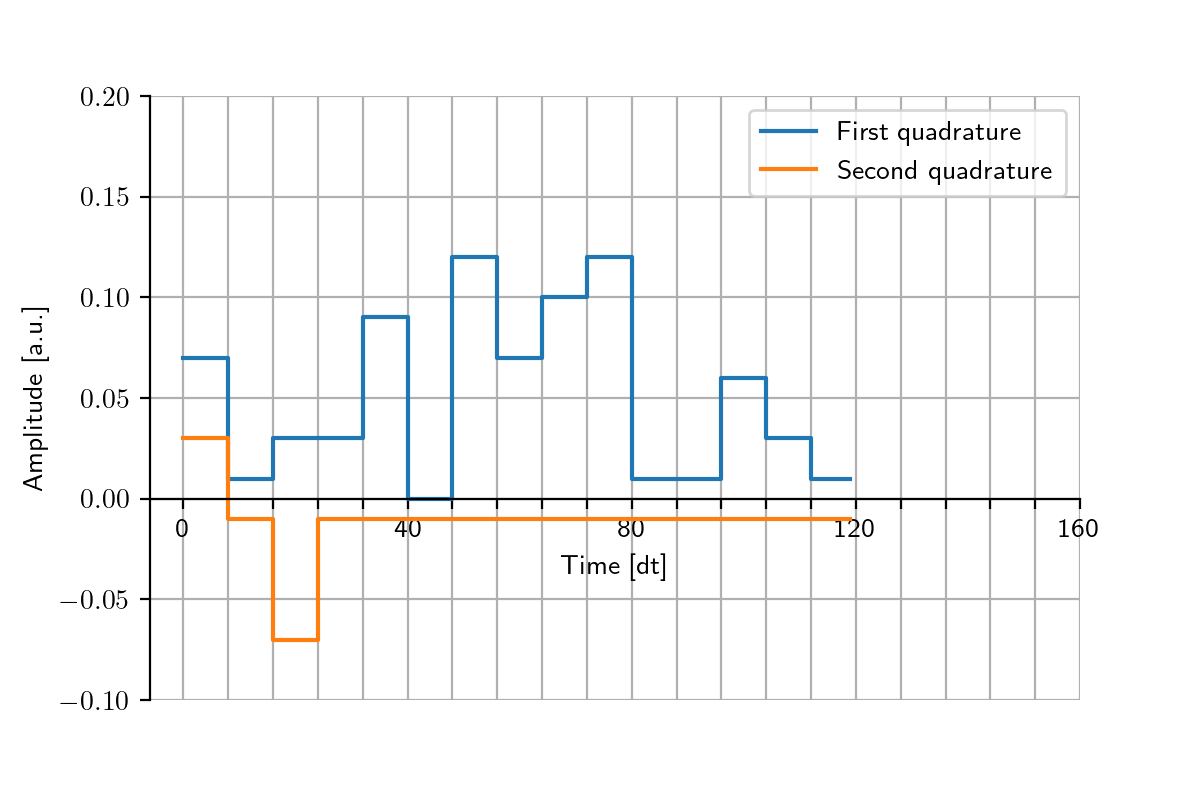}
        \vspace{-10mm}
        \caption{Optimized $X$ gate $1.33\times$ faster.}
        \label{fig:XresC}
    \end{subfigure}
    \hspace{-5mm}
    \begin{subfigure}[b]{0.49\textwidth}
        \includegraphics[clip,width=\textwidth]{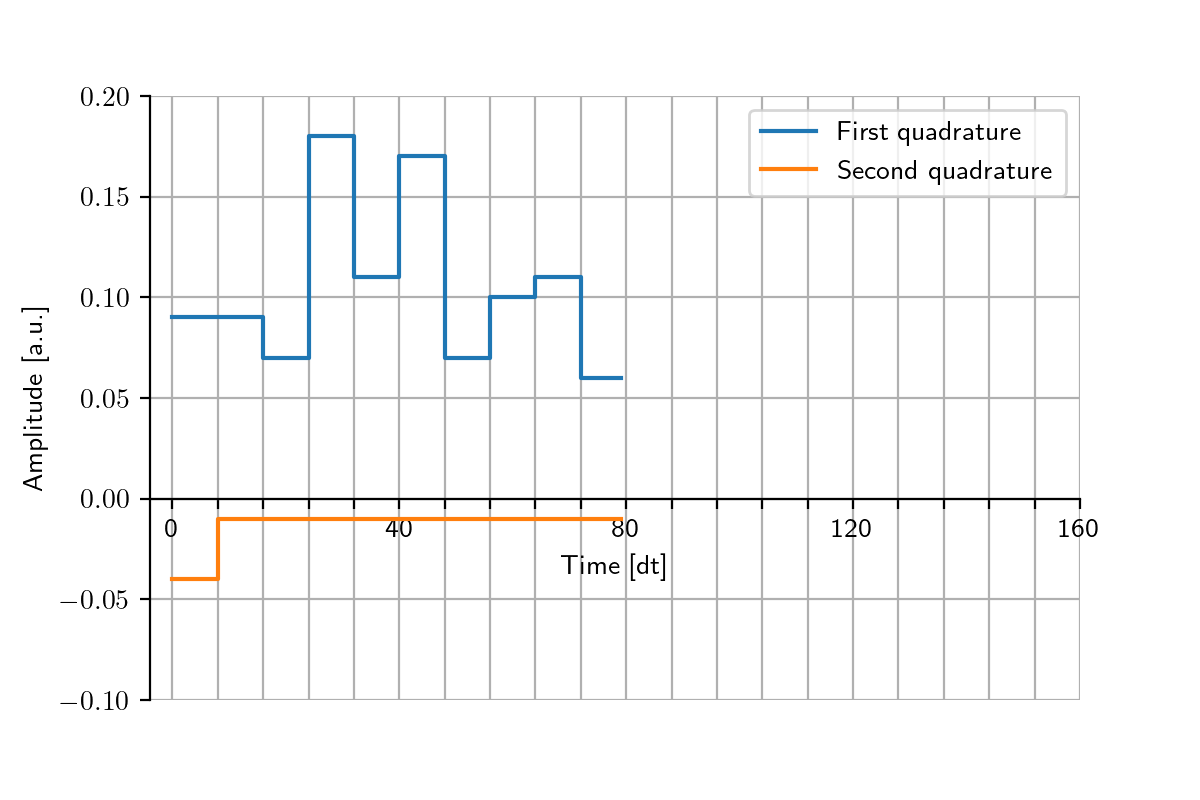}
        \vspace{-10mm}
        \caption{Optimized $X$ gate $2\times$ faster.}
        \label{fig:XresD}
    \end{subfigure}
    \\
    \centering
    \begin{subfigure}[b]{0.49\textwidth}
        \includegraphics[clip,width=\textwidth]{Figures/X_Default_Gate.png}
        \vspace{-10mm}
        \caption{DRAG $X$ gate calibrated after 30 days.}
        \label{fig:XresE}
    \end{subfigure}
    \hspace{-5mm}
    \begin{subfigure}[b]{0.49\textwidth}
        \includegraphics[clip,width=\textwidth]{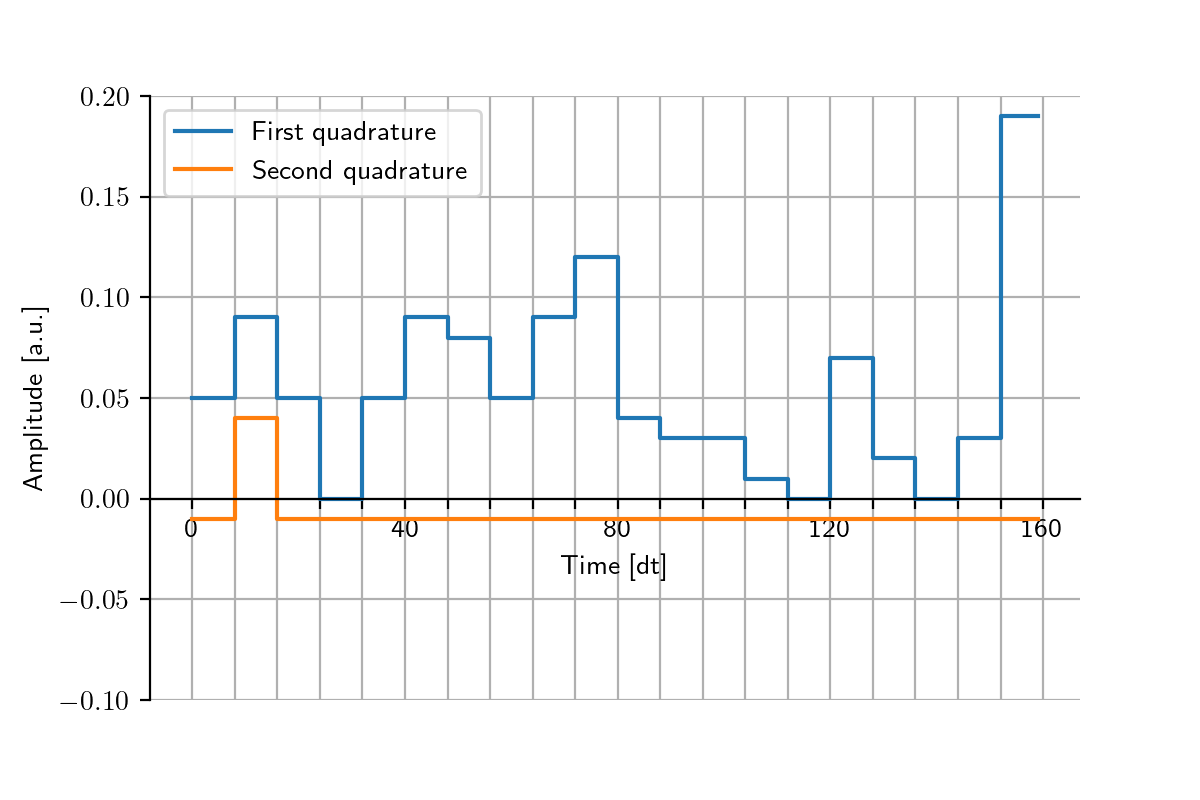}
        \vspace{-10mm}
        \caption{Optimized $X$ gate calibrated after 30 days.}
        \label{fig:XresF}
    \end{subfigure}
    \\
    \centering
    \vspace{2mm}
    \begin{subfigure}[b]{0.65\textwidth}
        \includegraphics[clip,width=\textwidth]{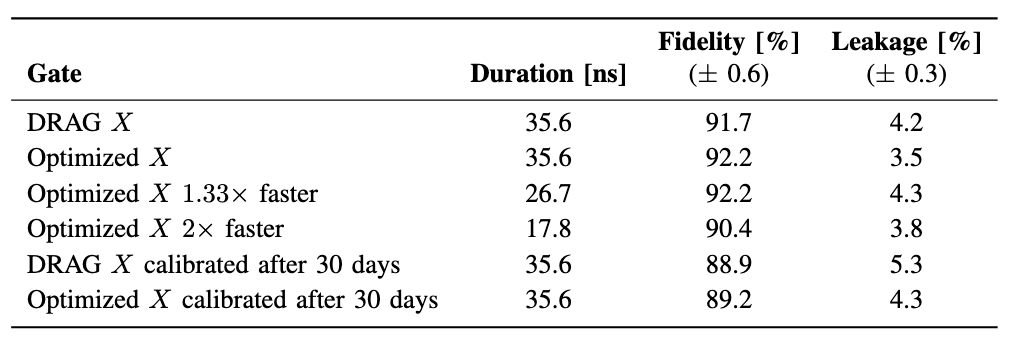}
        \vspace{-5mm}
        \caption{Summary of fidelities and leakage rates.}
        \label{fig:XresTab}
    \end{subfigure}
    \vspace{3mm}
    \caption{Plots showing default and optimized control pulses for single qubit gates on the IBM Lima quantum computer and their respective state fidelities and leakage rates.
    The time unit on the $x$-axis is in units of $dt = 0.222222$ ns, a device-dependent parameter specifying the maximum sampling rate of the waveform generator.
    Fig. (a) is a DRAG pulse for an $X$ gate.
    Fig. (b) is a pulse for an $X$ gate with duration 35.6 ns learned by our deep RL algorithm.
    Fig. (c) is a pulse for an $X$ gate with duration 26.7 ns learned by our deep RL algorithm.
    Fig. (d) is a pulse for an $X$ gate with duration 17.8 ns learned by our deep RL algorithm.
    Fig. (e) is a pulse for an $X$ gate calibrated by our deep RL algorithm 30 days after training.
    Fig. (f) is a DRAG pulse for an $X$ gate calibrated 30 days after training.
    Fig. (g) is table summarizing the state fidelities (\ref{eq:fid}) and leakage rates (\ref{eq:leak}) achieved by these gates.
    }
    \label{fig:Xres}
\end{figure*}

\begin{figure*}
    \centering
    \begin{subfigure}[b]{0.49\textwidth}
        \includegraphics[clip,width=\textwidth]{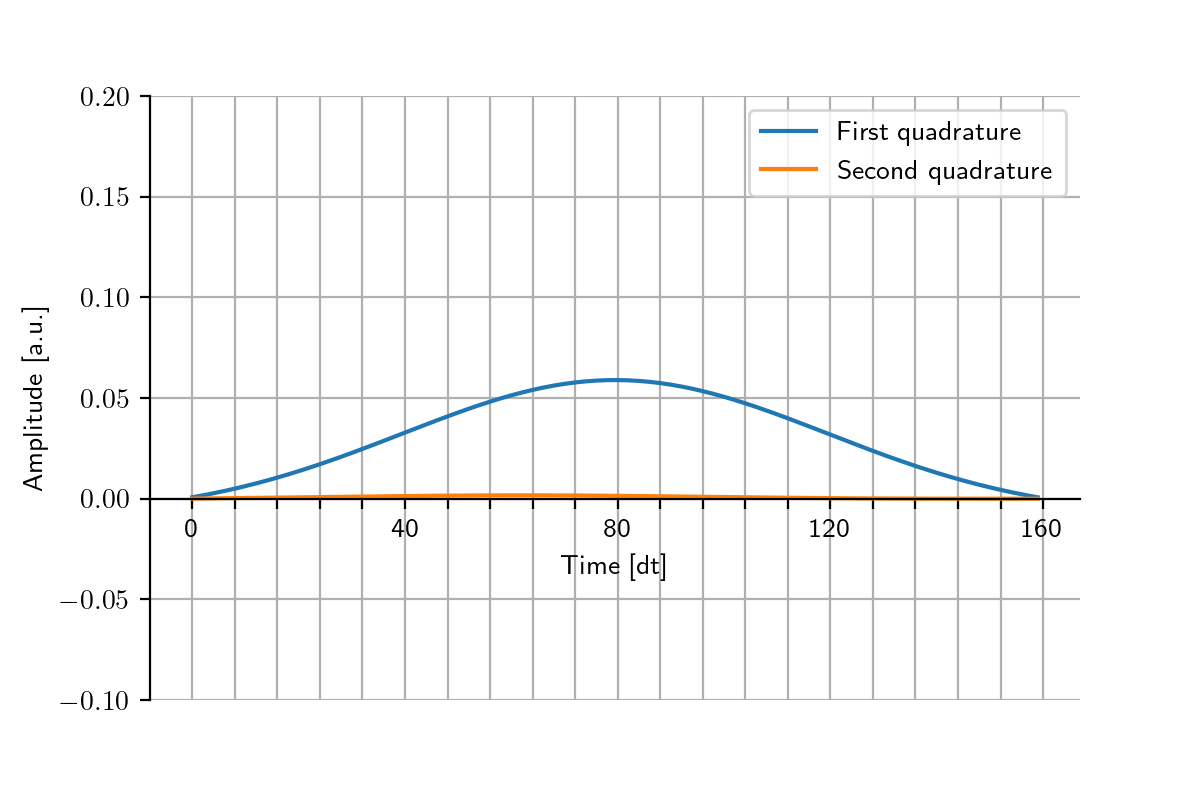}
        \vspace{-10mm}
        \caption{DRAG $\sqrt{X}$ gate.}
        \label{fig:SXresA}
    \end{subfigure}
    \hspace{-5mm}
    \begin{subfigure}[b]{0.49\textwidth}
        \includegraphics[clip,width=\textwidth]{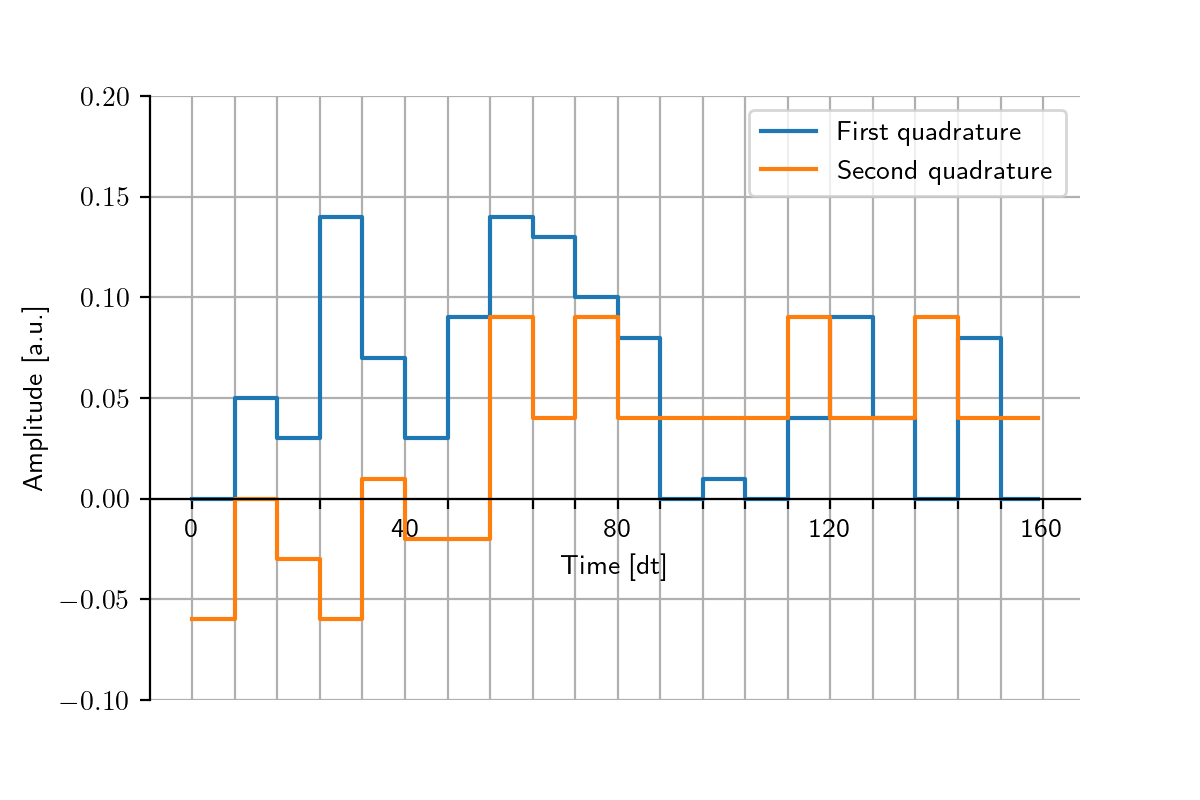}
        \vspace{-10mm}
        \caption{Optimized $\sqrt{X}$ gate.}
        \label{fig:SXresB}
    \end{subfigure}
    \vspace{-4mm}
    \\
    \centering
    \begin{subfigure}[b]{0.49\textwidth}
        \includegraphics[clip,width=\textwidth]{Figures/X_Optimal_Gate_15_seg.png}
        \vspace{-10mm}
        \caption{Optimized $\sqrt{X}$ gate $1.33\times$ faster.}
        \label{fig:SXresC}
    \end{subfigure}
    \hspace{-5mm}
    \begin{subfigure}[b]{0.49\textwidth}
        \includegraphics[clip,width=\textwidth]{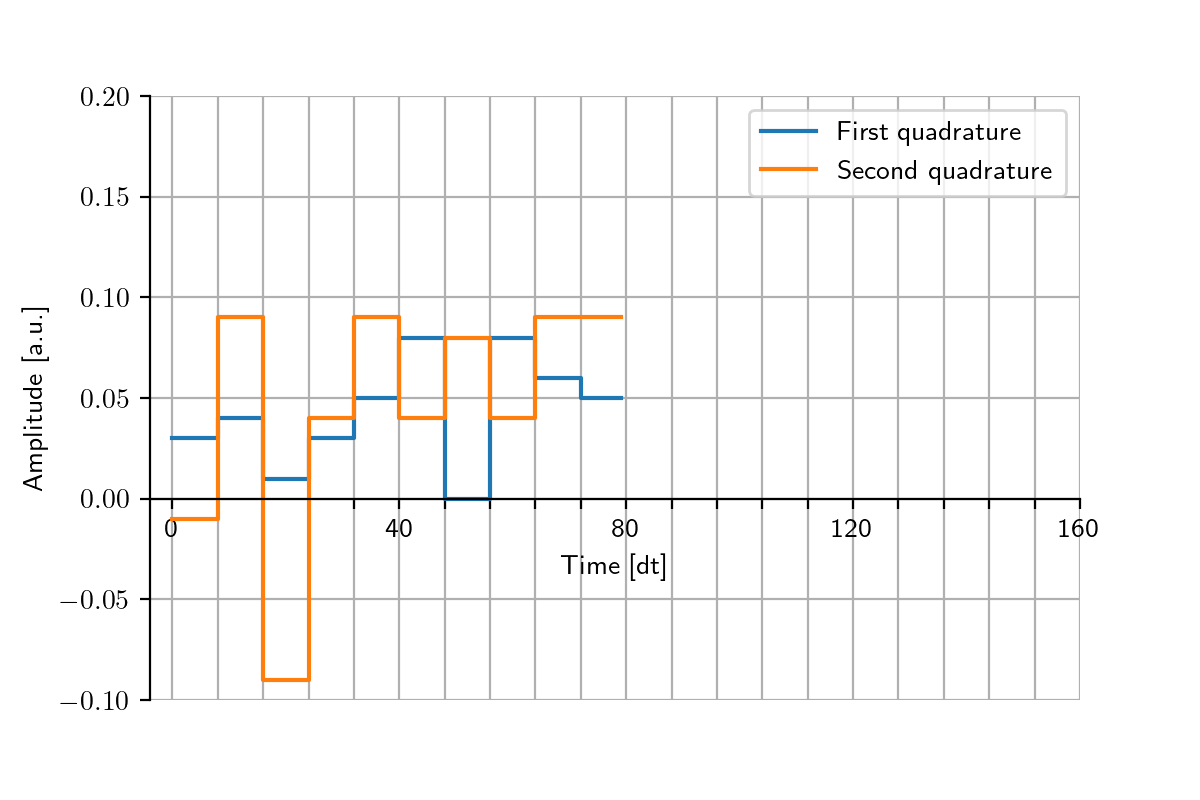}
        \vspace{-10mm}
        \caption{Optimized $\sqrt{X}$ gate $2\times$ faster.}
        \label{fig:SXresD}
    \end{subfigure}
    \vspace{2mm}
    \\
    \centering
    \begin{subfigure}[b]{0.6\textwidth}
        \includegraphics[clip,width=\textwidth]{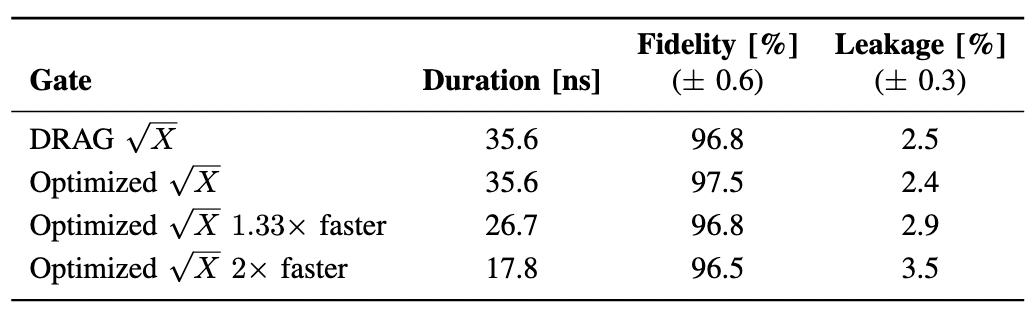}
        \vspace{-6mm}
        \caption{Summary of fidelities and leakage rates.}
        \label{fig:SXresTab}
    \end{subfigure}
    \vspace{3mm}
    \caption{Plots showing default and optimized control pulses for single qubit gates on the IBM Lima quantum computer and their respective state fidelities and leakage rates.
    The time unit on the $x$-axis is in units of $dt = 0.222222$ ns, a device-dependent parameter specifying the maximum sampling rate of the waveform generator.
    Fig. (a) is a DRAG pulse for an $\sqrt{X}$ gate.
    Fig. (b) is a pulse for an $\sqrt{X}$ gate with duration 35.6 ns learned by our deep RL algorithm.
    Fig. (c) is a pulse for an $\sqrt{X}$ gate with duration 26.7 ns learned by our deep RL algorithm.
    Fig. (d) is a pulse for an $\sqrt{X}$ gate with duration 17.8 ns learned by our deep RL algorithm.
    Fig. (e) is table summarizing the state fidelities (\ref{eq:fid}) and leakage rates (\ref{eq:leak}) achieved by these gates.
    }
    \label{fig:SXres}
\end{figure*}

\section{Experimental Results}
\label{sec:exp}
We conduct a proof-of-concept by designing the $X$ and $\sqrt{X}$ gates on the IBM Lima quantum computer using the Qiskit Pulse library~\cite{Qiskit, McKay2018}.
To assess the success of a gate $U$, we measure the leakage rate $\Ell$ defined in (\ref{eq:leak}) and the state fidelity
\begin{equation}
    \label{eq:fid}
    \F = \left|\bra{\psi_T}U\ket{0}\right|^2
\end{equation}
where $\ket{\psi_T}$ is the target state.
We benchmark our fast quantum gates against DRAG gates, which are the industry standard used on many quantum computers including those produced by IBM.  
The trained agent can create gates of any number of segments from 1 to $N_\textup{seg}$.
We test gates of length 20 segments ($t_g \approx 35.6$ ns) -- the same length as the DRAG gate --, 15 segments ($t_g \approx 26.7$ ns), and 10 segments ($t_g \approx 17.8$ ns).
After just 200 training iterations, the faster gates begin to match the performance of the default gate, with fidelities and leakage rates the same or only slightly worse.
Our full-length optimized gates begin to achieve higher state fidelities and lower leakage rates than the calibrated DRAG pulses.
The optimized $X$ and $\sqrt{X}$ gates are shown in Figures \ref{fig:Xres} and \ref{fig:SXres} respectively alongside the corresponding DRAG pulses.
The fidelities and leakage rates for each gate are summarized in Figures \ref{fig:XresTab} and \ref{fig:SXresTab}.

We also test the robustness of our agent over time.
Figure \ref{fig:XresF} shows a new optimized $X$ gate created 30 days after the agent was trained.
Again, it has slightly higher fidelity and lower leakage than the calibrated DRAG pulse.
This shows that our agent does not require any additional training after 30 days.
After an initial training period, our algorithm can be used to efficiently calibrate gates on superconducting qubits.

Due to limited access to hardware, we do not train our agent beyond 200 iterations nor do we optimize the learning rate, neural network structure, number of segments, or other hyperparameters.
Our algorithm has not fully converged after the first 200 training iterations.
We expect to achieve more significant advantage after optimizing the algorithm hyperparameters including using more training iterations.

\section{Generalization to two-qubit gates}
\label{sec:gen}
A small modification to state space and reward function generalizes our algorithm to the design of two-qubit gates.
In superconducting transmon qubits, two-qubit interactions are generated via a cross-resonance ($CR$) pulse (i.e. driving the control qubit at the resonant frequency of the target qubit)~\cite{Paraoanu2006}. 
The $CR$ pulse corresponds to the gate $Z \otimes X$, an $X$ rotation on the target qubit with the direction dependent on the state of the control qubit. 
The industry standard $CR$ gate is a rounded square pulse with Gaussian rise and a derivative-pulse correction on the second quadrature~\cite{Sheldon2016}.
To avoid spurious cross-talk, a simultaneous cancellation tone is applied to the target qubit. 
Baum et al. used RL to design an improved $CR$ gate which did not require this extra pulse~\cite{Baum2021}.
We take the same strategy, so our agent can learn an improved $CR$ pulse shape using the action spaces $\bb{U}^x$ and $\bb{U}^y$ from the single qubit case.
The state $\tilde{s}_j^x = (\langle I_j^1\rangle, \langle Q_j^1\rangle, \langle I_j^2\rangle, \langle Q_j^2\rangle, k)$ contains the average readout signal from each qubit.
We use the reward 
\begin{align}
    \label{eq:rx2}
    &\tilde{r}_j^x(\tilde{s}_{j+1}^x, c_T^1, c_T^2) = \\
    &\frac{1}{2}\left(r_j^x(\langle I_{j+1}^1\rangle, \langle Q_{j+1}^1\rangle, k, c_T^1) + r_j^x(\langle I_{j+1}^2\rangle, \langle Q_{j+1}^2\rangle, k, c_T^2)\right) \notag
\end{align}
where $r_j^x$ is defined in (\ref{eq:rx}) and $c_T^1$, $c_T^2$ are the expected measurement locations for the first and second qubit respectively.
For the second control quadrature, the leakage population is summed over both qubits to give the same state and reward as the single qubit case.
The $CR$ gate completes a universal gateset $\left\{CR, R_Z(\theta), X, \sqrt{X}\right\}$ when combined with arbitrary virtual $Z$ rotations~\cite{McKay2017} and the single qubit gates shown in our proof-of-concept.

\section{Conclusion}
\label{sec:conc}
We have created a DRL algorithm for fast quantum gate design using real-time feedback based on the readout signal from noisy hardware.
RL is model free, as opposed to other gate synthesis strategies which require precise models that cannot capture the stochastic dynamics of the qubit.
Our dual-agent architecture allows us to target competing goals of decreasing leakage and creating faster gates to reduce decoherence.
Our proposed algorithm reduces the impact of measurement classification errors by training directly on the readout signal from probing the qubit.
The low measurement overhead means that our algorithm can be trained on hardware, rather than a numerical simulation.

We carried out a proof-of-concept with $X$ and $\sqrt{X}$ gates of different durations on IBM’s hardware.
Our agent proposed novel control pulses which are two times faster than industry standard DRAG gates and match their performance in terms of fidelity and leakage after just 200 training iterations.
Our agent also created gates of the same duration as DRAG which offer slight improvements in fidelity and leakage.
We also showed that our trained agent is robust over time.

So far, we had limited access to hardware and only ran the algorithm with one specific set of hyperparameters.
As is well known, the performance of RL is greatly improved with fine tuning of hyperparameters~\cite{Feurer2019}.
We expect optimization of the algorithm hyperparameters to lead to much more significant advantage.

Our proof-of-concept was carried out on transmon qubits; however, our proposed DRL algorithm is general and can be used on other quantum hardware with adjustments to the state and action spaces (e.g. photon counts and laser pulses for trapped ion quantum computing~\cite{Bruzewicz2019}).
The improved gate operations created by our DRL algorithm for fast quantum gate design open the way for more extensive applications on near term and future quantum devices with reduced decoherence and leakage which cannot be fixed by most error correction algorithms.

\appendices
\section{Transmon hardware}
\label{sec:hardware}
We apply our deep RL algorithm for fast quantum gate design to a transmon qubit, formed of the lowest two energy levels in the quantized spectrum of a superconducting circuit~\cite{Koch2007}. 
The transmon qubit is controlled via a capacitively coupled drive line.
Quantum circuit analysis leads to the transmon Hamiltonian~\cite{Krantz2019}
\begin{equation}
    \label{eq:transHamCtrl}
    \h = \omega_qa^\dg a + \frac{\alpha}{2}a^\dg a^\dg aa - ic(t)(a^\dg - a)
\end{equation}
where $a$, $a^\dg$ are the creation and annihilation operators, $\omega_q$ is the qubit frequency, $\alpha$ is the anharmonicity and $c(t)$ is the voltage envelope introduced in (\ref{eq:drive}).

The transmon qubit is dispersively coupled to a superconducting resonator for measurement.
The qubit is probed with a microwave signal, which scatters off the resonator and gets amplified.
The resultant observation is comprised of time traces of the in-phase $I$ and quadrature $Q$ components of the digitized signal.
The state of the qubit can be inferred from the location of the signal in the $(I, Q)$-plane.
On noisy hardware, the locations often overlap in the $(I, Q)$-plane and a decision maker such as a linear discriminant analyzer~\cite{Magesan2015} is necessary to predict the state (see Figure \ref{fig:stateData}).

\section{Pre-training}
\label{sec:pretrain}
In this section, we describe pre-training our agent.
First, we discretize a calibrated DRAG pulse into an $N_\textup{seg}$ PWC pulse where the amplitude of each segment is the nearest action from $\bb{U}^x$ and $\bb{U}^y$ for the first and second quadratures respectively.
For each $k$, we measure the average $(\langle I\rangle, \langle Q\rangle)$ signal and leakage $\Ell$ after the first $k+1$ segments of the waveform.
Each agent receives a reward based on the mean squared error (MSE) between the policy output by the neural network given the input state and a Gaussian distribution over the respective action space centered at the desired amplitude.
The pre-training is performed for several iterations over the full pulse.
The pre-training can be executed efficiently because the rewards and actions are not based on the state of the system so the circuits can all be run before updating the network parameters.
At the end of the pre-training, the agents favour a DRAG pulse.
While the agents subsequently learn novel pulse shapes, the exploration begins in a near-optimal region of the solution space. 


\bibliographystyle{IEEEtran}
\bibliography{biblio}
\end{document}